\documentclass[journal,10pt]{IEEEtran}
\IEEEoverridecommandlockouts
\usepackage{cite}
\usepackage{amsmath,amssymb,amsfonts}
\usepackage{algorithmic}
\usepackage{graphicx}
\usepackage{textcomp}
\usepackage{xcolor}
\usepackage{amsmath}
\usepackage{cases}
\usepackage{algorithm}  
\usepackage{booktabs}
\usepackage{color,soul}
\usepackage{subfigure}
\ifCLASSOPTIONcompsoc 
\usepackage[caption=false,font=normalsize,labelfon t=sf,textfont=sf]{subfig} 
\else 
\usepackage[caption=false,font=footnotesize]{subfi g} \fi

\makeatletter

\newcommand{\Rmnum}[1]{\expandafter\@slowromancap\romannumeral #1@}
\newcommand{\removelatexerror}{\let\@latex@error\@gobble}
\makeatother
\def\BibTeX{{\rm B\kern-.05em{\sc i\kern-.025em b}\kern-.08em
    T\kern-.1667em\lower.7ex\hbox{E}\kern-.125emX}}
\begin{document}

%
\title{Min-Max Latency Optimization Based on Sensed Position State Information in Internet of Vehicles\\
}

\author{Pengzun Gao, Long Zhao, \IEEEmembership{Member, IEEE,} Kan Zheng, \IEEEmembership{Senior Member, IEEE} and Pingzhi Fan, \IEEEmembership{Fellow, IEEE}
\IEEEcompsocitemizethanks{\IEEEcompsocthanksitem P. Gao, L. Zhao and K. Zheng are with Intelligent Computing and Communications  Lab, Beijing University of Posts and Telecommunications (BUPT), Beijing, 100876, China (e-mail: gaopengzun@bupt.edu.cn; z-long@bupt.edu.cn; zkan@bupt.edu.cn).}
\IEEEcompsocitemizethanks{\IEEEcompsocthanksitem P. Fan is with the Information Coding and Transmission Key Lab of Sichuan Province, CSNMT Int. Coop. Res. Centre (MoST), Southwest Jiaotong University (SWJTU), Sichuan, 611756, China (e-mail: pzfan@swjtu.edu.cn).}
}

\maketitle

\begin{abstract}
The dual-function radar communication (DFRC) is an essential technology in Internet of Vehicles (IoV). Consider that the road-side unit (RSU) employs the DFRC signals to sense the vehicles’ position state information (PSI), and communicates with the vehicles based on PSI. The objective of this paper is to minimize the maximum communication delay among all vehicles by considering the estimation accuracy constraint of the vehicles' PSI and the transmit power constraint of RSU. By leveraging convex optimization theory, two iterative power allocation algorithms are proposed with different complexities and applicable scenarios. Simulation results indicate that the proposed power allocation algorithm converges and can significantly reduce the maximum transmit delay among vehicles compared with other schemes.
\end{abstract}

\IEEEpeerreviewmaketitle

\section{Introduction}

Internet of Vehicles (IoV) has recently attracted the attention of many researches as a typical application scenario of ultra-reliable low-latency communications (uRLLC) in next generation (NG) communications~\cite{a1}. The aim of IoV is to improve the transmit rate and simultaneously reduce the transmit delay~\cite{a2,a3,a4,a34}. The main methods to achieve this aim can be classified into two categories: traditional methods such as convex optimization~\cite{a33,a6,a7} and intelligent methods such as machine learning~\cite{a8,a9,a10}.

In order to maximize the downlink beam gain, the vehicle needs to upload its position state information (PSI) periodically, which results in significant uplink overhead. However, by combining the radar and communication~\cite{a31,a32,a30}, the road side unit (RSU) can sense vehicles' PSI and simultaneously transmit information to vehicles, which effectively cuts the uplink system overhead. The current literature shows that two different ways can be employed to combine the radar and communication~\cite{a25}. One way is the primary collaboration between radar system and communication system~\cite{a26}. However, with the increasing communication frequency, the bands of radar and communication may be overlapped, which may cause severe interference. Therefore, the other way is to integrate radar and communication functions into the same hardware and spectral resources, i.e., dual-function radar communication (DFRC)~\cite{a12}.

When the DFRC is employed for sensing and communication in IoV systems, the maximum rate optimization problem has been studied by taking into account the communication power constraints~\cite{a24}. While the sensing accuracy has been not taken into full account, which significantly deteriorates the communication performance. Moreover, the sensing accuracy optimization problem has been studied by considering the communication rate and power constraints, however ignoring the main metric of IoV, i.e., transmit delay~\cite{a14}. In this paper, in order to minimize the maximum communication delay among vehicles in the DFRC systems, the vehicles' PSI constraints as well as the transmit power constraint of RSU are taken into account. By problem analysis and transformation, two efficient algorithms are proposed with different computational complexities and applicable conditions. According to the simulation results, the proposed algorithms can achieve microsecond-level latency, meeting the latency requirements of NG communications~\cite{a29}.

\section{System Model and Problem Formulation}

\begin{figure}[!b]
	\centering
	\includegraphics[width=3.5in]{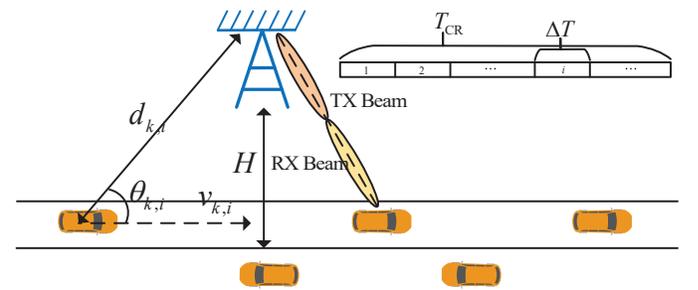}
	\caption{Illustration of DFRC system in IoV.}
	\label{1}
\end{figure}
As depicted in Fig. 1, a millimeter wave (mmWave) RSU equipping with $N_\text{t}$ transmit antennas and $N_\text{r}$ receive antennas transmits DFRC signals to $K$ $N_\text{v}$-antenna vehicles. Both the RSU and vehicles adopt uniform linear arrays (ULA), and the antenna spacing of which is assumed to be half-wavelength. Denote $T_\text{CR}$ as the maximum time duration of interest, and $T_\text{CR}$ is divided into several time-slots, i.e., $\Delta T$. Denote ${{\bf{x}}_i} = \left[\theta_{k,i},d_{k,i},v_{k,i},\beta_{k,i} \right] ^ \text{T}$ as the $k$th vehicle's PSI at the $i$th time-slot, where $\theta_{k,i}$, $d_{k,i}$, $v_{k,i}$ and $\beta_{k,i}$ are the angle of departure (AoD) of the signal from the RSU to the $k$th vehicle, the distance between the RSU and the $k$th vehicle, the velocity and the radar channel coefficient of the $k$th vehicle (contains the large-scale fading factor and the radar cross-section (RCS)), respectively. We assume that the vehicle's PSI remains constant during $\Delta T$, and relates between adjacent time-slots. Based on the estimated vehicles' PSI at previous time-slots and current time-slot, the transmit power at the current time-slot should be allocated in order to optimize the transmit delay among vehicles.

\subsection{Signal Model}
\subsubsection{Transmission signal model}
The RSU simultaneously transmits $K$ symbols to $K$ vehicles
, which can be written as

\begin{equation}
{{\bf{s}}_i}\left( t \right) = {\left[ {{s_{1,i}}\left( t \right),{s_{2,i}}\left( t \right), \cdots ,{s_{K,i}}\left( t \right)} \right]^{\rm{T}}} \in {\mathbb{C}^{K \times 1}},
\end{equation}
where ${\left| {{s_{k,i}}\left( t \right) } \right|} = 1$. 

Assuming that ${\mathbf{u}}_{k,i}$ represents the transmit precoder for the $k$th vehicle, the transmit signal ${\bf{f}}_{i}\left( t \right)$ of the RSU can be given by

\begin{equation}
{{\bf{f}}_{i}\left( t \right)} = \sum\limits_{k=1}^K {\mathbf{u}_{k,i}}{s_{k,i}}\left( t \right).
\end{equation}

%

\subsubsection{Received communication signal model}
Denote ${\phi _{k,i}}$ as the angle of arrival (AoA) of the signal from the RSU to the $k$th vehicle. And assuming that the ULA of the RSU is parallel to the straight road, we have ${\theta _{k,i}} = {\phi _{k,i}}$. Then, the transmit and receive steering vectors can be respectively written as~\cite{a28}
\begin{equation}
	{\bf{a}}\left( {{\theta _{k,i}}} \right) = {\sqrt {\frac{1}{{{N_{\text{t}}}}}} \left[ {1,{\operatorname{e} ^{ - j\pi \cos {\theta _{k,i}}}}, \cdots ,{\operatorname{e} ^{ - j\left( {{N_\text{t}} - 1} \right)\pi \cos {\theta _{k,i}}}}} \right]^{\text{T}}},
\end{equation}
\begin{equation}
	{\bf{v}}\left( {{\theta _{k,i}}} \right) = {\sqrt {\frac{1}{{{N_{\text{v}}}}}} \left[ {1,{\operatorname{e} ^{ - j\pi \cos {\theta _{k,i}}}}, \cdots ,{\operatorname{e} ^{ - j\left( {{N_\text{v}} - 1} \right)\pi \cos {\theta _{k,i}}}}} \right]^{\text{T}}}.
\end{equation}

Moreover, the Doppler Shift can be assumed as ${\varsigma _{k,i}} =  {{{v_{k,i}}\cos \left( {{\theta _{k,i}}} \right){f_\text{c}}} \mathord{\left/{\vphantom {{{v_{k,i}}\cos \left( {{\phi _{k,i}}} \right){f_\text{c}}} c}} \right.\kern-\nulldelimiterspace} c}$, where  $f_\text{c} \text{ and } c $ represent the carrier frequency and the speed of light, respectively.
Therefore, the channel matrix between the RSU and the $k$th vehicle can be expressed as
\begin{equation}
	{\bf{H}_\text{C}}\left( {{\theta _{k,i}}} \right)  = {\alpha _{k,i}}{{\rm{e}}^{j2\pi {\varsigma _{k,i}}t}}{\bf{v}}\left( {{\theta _{k,i}}} \right){{\bf{a}}^{\rm{H}}}\left( {{\theta _{k,i}}} \right), \label{channel c}
\end{equation}
where ${\alpha _{k,i}} = \tilde \alpha d_{k,i}^{ - 1}{\operatorname{e}^{j\frac{{2\pi {f_\text{c}}}}{c}{d_{k,i}}}}$ is the large-scale fading factor and $\tilde \alpha$ is a constant~\cite{a14}.

On one hand, we employ the matched filter (MF) precoder and detector, i.e., the transmit precoder ${{\bf{u}}_{k,i}} = {\bf{a}}\left( {{\hat \theta _{k,i\left| {i - 1} \right.}}} \right)$ and the detector ${{\bf{w}}_{k,i}} = {\bf{v}}\left( {{{\hat \theta }_{k,i\left| {i - 2} \right.}}} \right)$, where ${{\hat \theta _{k,i\left| {i - 1} \right.}}}$ (${{\hat \theta _{k,i\left| {i - 2} \right.}}}$) represents the one-slot (two-slots) prediction of angle parameter. On the other hand, by considering the channel hardening effect and asymptotic channel orthogonality of mMIMO system~\cite{a23}, i.e.,
\begin{equation}
\left| {{{\bf{a}}^{\rm{H}}}\left( \theta  \right){\bf{a}}\left( \phi  \right)} \right| \to 0,\forall \theta  \ne \phi ,{N_{\rm{t}}} \to \infty. \label{MIMO}
\end{equation}
The received communication signal at the $k$th vehicle can be then written as
\begin{align}
	r_{k,i}^{\text{C}}\left( t \right) =\kappa \sqrt {p_{k,i}}{\mathbf{w}}_{k,i}^{\text{H}} {\bf{H}_\text{C}}\left( {{\theta _{k,i}}} \right){{\mathbf{u}}_{k,i}}s_{k,i}\left( t \right) + {z^{\text{C}}}\left( t \right), \label{e40}
\end{align}
where $\kappa = \sqrt{{N_\text{t}}{N_\text{v}}}$ is the array gain factor, $p_{k,i}$ represents the transmit power for the $k$th vehicle, and ${z^{\text{C}}}\left( t \right) \sim \mathbb{C}\mathbb{N}\left( {0,\sigma _{\text{C}}^2} \right)$ is the complex Gaussian noise~\cite{a14,a16}.

\subsubsection{Received radar signal model}
Denote ${\bf{b}}\left( {{\theta _{k,i}}} \right) \in {{\mathbb{C}}^{{N_\text{r}} \times 1}}$ as the receive steering vector, which is similar to ${\bf{a}}\left( {{\theta _{k,i}}} \right)$, and the round trip channel matrix between the RSU and the $k$th vehicle can be expressed as
\begin{equation}
	{\bf{H}_\text{R}}\left( {{\theta _{k,i}}} \right) = {\beta _{k,i}}{{\rm{e}}^{j2\pi {\mu _{k,i}}t}}{\bf{b}}\left( {{\theta _{k,i}}} \right){{\bf{a}}^{\rm{H}}}\left( {{\theta _{k,i}}} \right), \label{channel r}
\end{equation}
where ${\mu _{k,i}} = {{2{v_{k,i}}\cos \left( {{\theta _{k,i}}} \right){f_\text{c}}} \mathord{\left/{\vphantom {{2{v_{k,i}}\cos \left( {{\theta _{k,i}}} \right){f_c}} c}} \right.\kern-\nulldelimiterspace} c}$ is the Doppler Shift.

According to (\ref{MIMO}) and (\ref{channel r}), the received radar signal of the RSU reflected by the $k$th vehicle can be given by 
\begin{align}
	{\bf{r}}_{k,i}^\text{R}\left( t \right) = \tilde \kappa \sqrt {{p_{k,i}}}{\bf{H}_\text{R}}\left( {{\theta _{k,i}}} \right){{\bf{u}}_{k,i}}s_{k,i}\left( {t - {\tau _{k,i}}} \right) + {{\bf{z}}_{k,i}}\left( t \right),
\end{align}
where $\tilde \kappa = \sqrt{{N_\text{t}}{N_\text{r}}}$ is the array gain factor, ${\tau _{k,i}} = {{2{d_{k,i}}} \mathord{\left/{\vphantom {{2{d_{k,i}}} c}} \right.\kern-\nulldelimiterspace} c}$ represents the delay between echo signal and transmit signal, and ${{\mathbf{z}}_{k,i}}\left( t \right) \sim \mathbb{C}\mathbb{N}\left( {0,\sigma _{\text{R}}^2{{\mathbf{1}}_{{N_{\text{r}}}}}} \right) $ denotes the complex Gaussian noise~\cite{a14,a16}.


\subsection{Problem Formulation}
According to (\ref{e40}) and assuming that the system bandwidth is $B$, the transmit rate of the $k$th vehicle can be given by
\begin{equation}
{R_{k,i}} = B \log _2 \left[ {1 +  \frac{{p_{k,i}\kappa^2}}{{\sigma _\text{c}^2} } {{{\left| {{\bf{w}}_{k,i}^{\text{H}}{\bf{H}_\text{C}}\left( {{\theta _{k,i}}} \right){{\bf{u}}_{k,i}}} \right|}^2}}    } \right], \label{e300}
\end{equation}
therefore, the transmit delay of the $k$th vehicle can be expressed as
\begin{equation}
{T_{k,i}} = \frac{D_{k,i}}{R_{k,i}},\label{e30}
\end{equation}
where $D_{k,i}$ is the size of the delay-sensitive information for the $k$th vehicle~\cite{a21}.

Limited by the total transmit power $P_{\text{M}}$ of the RSU and the maximum tolerant delay ${T_\text{d}}$, the objective of this paper is to minimize the maximum transmit delay among $K$ vehicles in each time-slot while guaranteeing the estimation accuracy of vehicles' PSI, i.e., the optimization problem can be formulated as
\begin{align}
	{\cal P}{1}:\;  &\mathop {\text{min} }\limits_{{p_{k,i}}} {\text{ }}\mathop {{\text{max}}}\limits_k {\text{ }}{T_{k,i}} \label{e2}\\ 
	{\text{s}}{\text{.t}}{\text{. }} &{\text{PCRB}}\left( {{\theta _{k,i}}} \right) \leqslant {\xi _\theta },\forall k, \label{e3} \\ 
	&{\text{PCRB}}\left( {{d_{k,i}}} \right) \leqslant {\xi _d},\forall k, \label{e4}  \\ 
	&{T_{k,i}} \leqslant \min \left\{ {{\Delta T},{T_\text{d}}} \right\},\forall k, \label{e5} \\ 
	&\sum\limits_{k=1}^K {{p_{k,i}}}  \leqslant {P_{{\text{M}}}} \label{e6},  
\end{align} 
where ${\text{PCRB}}\left( {{\theta _{k,i}}} \right)$ and ${\text{PCRB}}\left( {{d_{k,i}}} \right)$ are the Cramer-Rao Bounds with Prior knowledge (PCRBs) of the angle and distance between the $k$th vehicle and RSU, respectively; $\xi _\theta $ and $\xi _d$ are the thresholds for ${\text{PCRB}}\left( {{\theta _{k,i}}} \right)$ and ${\text{PCRB}}\left( {{d_{k,i}}} \right)$, respectively. Generally, it is reasonable to only consider ${\text{PCRB}}\left( {{\theta _{k,i}}} \right)$ and ${\text{PCRB}}\left( {{d_{k,i}}} \right)$ as PSI constraints, because the velocity and RCS of the $k$th vehicle can be calculated by $\theta _{k,i}$ and $d_{k,i}$.

\section{Power Allocation for Min-max Latency}

\subsection{Problem Analysis}

PCRB is usually used to measure the positioning accuracy of radar systems in IoV, which is defined as the reciprocal of the Fisher matrix $\mathbf{J} = {{\mathbf{J}}_\text{o}}{+ }{{\mathbf{J}}_\text{p}}$~\cite{a17,a18}, where ${\mathbf{J}}_\text{o}$ and ${{\mathbf{J}}_\text{p}} $ represent the observed Fisher information and the prior Fisher information, respectively~\cite{a22}.
Therefore, in light of~\cite{a14}, assuming that $\mathbf{\Lambda }$ represents the diagonal matrix formed by the eigenvalues of ${\mathbf{J}}_{\text{p}}^{{{ - 1} \mathord{\left/
				{\vphantom {{ - 1} 2}} \right.
				\kern-\nulldelimiterspace} 2}}{{\mathbf{J}}_{\text{o}}}{\mathbf{J}}_{\text{p}}^{ - {{\text{H}} \mathord{\left/
				{\vphantom {{\text{H}} 2}} \right.
				\kern-\nulldelimiterspace} 2}}$, the PCRB matrix can be given by
	\begin{align}
		\mathbf{C} = {{\mathbf{J}}_\text{p}}{\left( {{p_{k,i}}{\mathbf{\Lambda }} + {\mathbf{I}}} \right)^{ - 1}}{{{\mathbf{J}}_{\text{p}}^{{\text{H}}}}} = \mathbf{J}^{-1} \in {\mathbb{C}^{4 \times 4}}.
	\end{align}
Whereas, the first two elements on the diagonal of PCRB matrix represent the PCRBs for the angle and distance, which can be respectively expressed as
	\begin{align}
		&{\text{PCRB}}\left( \theta_{k,i}  \right) = {c_{11}} = \sum\limits_{m = 1}^4 {\frac{{{{\left| {{{ b}_{1m}}} \right|}^2}}}{{{p_{k,i}}{\lambda _{m,k,i}} + 1}}} ,  \label{e7}\\
		&{\text{PCRB}}\left( d_{k,i} \right) = {c_{22}} = \sum\limits_{m = 1}^4 {\frac{{{{\left| {{{ b}_{2m}}} \right|}^2}}}{{{p_{k,i}}{\lambda _{m,k,i}} + 1}}}\label{e8},   
	\end{align}
	where $c_{ij}$ and ${ b_{ij}}$ denote the $(i,j)$ elements of $\mathbf{C}$ and ${{\mathbf{J}}_\text{p}} $, respectively; ${\lambda _{m,k,i}}\left(m = 1, 2,3 ,4\right)$ are the eigenvalues of $\mathbf{\Lambda }$.

\subsection{Problem Transformation}
Based on (\ref{e7}) and (\ref{e8}), the formulated problem ${\cal P}{1}$ can be transformed into
\begin{align}
	{\cal P}{2}:\; &\mathop {\text{min} }\limits_{{p_{k,i}}} {\text{ }}\mathop {{\text{max}}}\limits_k {\text{ }}{T_{k,i}} \label{e11}\\ 
	{\text{s}}{\text{.t}}{\text{. }} & \sum\limits_{m = 1}^4 {\frac{{{{\left| {{{b}_{1m}}} \right|}^2}}}{{{p_{k,i}}{\lambda _{m,k,i}} + 1}}}\leqslant {\xi _\theta },\forall k, \label{e12}\\ 
	& \sum\limits_{m = 1}^4 {\frac{{{{\left| {{{ b}_{2m}}} \right|}^2}}}{{{p_{k,i}}{\lambda _{m,k,i}} + 1}}}\leqslant {\xi _d },\forall k,  \label{e13}\\ 
	&{T_{k,i}} \leqslant \min \left\{ {\Delta T,{T_\text{d}}} \right\},\forall k,\label{e14}\\ 
	&\sum\limits_{k=1}^K {{p_{k,i}}}  \leqslant {P_{{\text{M}}}}.\label{e15}
\end{align} 
Because both the ${\text{PCRB}}\left( \theta_{k,i}  \right)$ and ${\text{PCRB}}\left( d_{k,i}  \right)$ are monotonically
decreasing functions of $p_{k,i}$, the lower bound $p_\text{m}$ exists based on (\ref{e12}) and (\ref{e13}). However, it is hard to obtain the closed-form of $p_\text{m}$. Therefore, to solve the above problem, we first relax the constrains (\ref{e12}) and (\ref{e13}) into 
\begin{align}
	&\sum\limits_{m = 1}^4 {\frac{{{{\left| {{{ b}_{1m}}} \right|}^2}}}{{{p_{k,i}}{\lambda _{m,k,i}}}}}\leqslant {\xi _\theta },\forall k, \label{e303}\\ 
	&\sum\limits_{m = 1}^4 {\frac{{{{\left| {{{ b}_{2m}}} \right|}^2}}}{{{p_{k,i}}{\lambda _{m,k,i}}}}}\leqslant {\xi _d },\forall k. \label{e304}
\end{align}
To simplify the expressions, we define 
\begin{align}
{{\cal A}_{k,i}} &= \left( {{{{D_{k,i}}} \mathord{\left/
			{\vphantom {{{D_{k,i}}} B}} \right.
			\kern-\nulldelimiterspace} B}} \right)\ln 2,\text{ and}\label{aa2}\\
{{\cal B}_{k,i}} &= {{{{\left| {{\kappa }{\bf{w}}_{k,i}^{\rm{H}}{{\bf{H}}_{\text{C}} \left({\theta _{k,i}}\right)}{{\bf{u}}_{k,i}}} \right|}^2}} \mathord{\left/
		{\vphantom {{{{\left| {{\alpha _{k,i}}{\bf{w}}_{k,i}^{\rm{H}}{\bf{v}}\left( {{\phi _{k,i}}} \right){{\bf{a}}^{\rm{H}}}\left( {{\theta _{k,i}}} \right){{\bf{u}}_{k,i}}} \right|}^2}} {\sigma _{\rm{C}}^2}}} \right. \kern-\nulldelimiterspace} {\sigma _{\rm{C}}^2}}.
\end{align}
Then, the optimization objective can be rewritten into
\begin{equation}
T_{k,i} = {{{{\cal A}_{k,i}}} \mathord{\left/
		{\vphantom {{{{\cal A}_{k,i}}} {\ln \left( {1 + {{\cal B}_{k,i}}{p_{k,i}}} \right)}}} \right.
		\kern-\nulldelimiterspace} {\ln \left( {1 + {{\cal B}_{k,i}}{p_{k,i}}} \right)}},\label{aa3}
\end{equation}
and the optimization problem 
${\cal P}2$ can be further simplified into          
\begin{align}
	{\cal P}{3}:\; &\mathop {\text{min} }\limits_{{p_{k,i}}} {\text{ }}\mathop {{\text{max}}}\limits_k {\text{ }}\frac{{{\cal A}_{k,i}}}{{\ln \left( {1 + {{\cal B}_{k,i}}{p_{k,i}}} \right)}} \label{e16}\\ 
	{\text{s}}{\text{.t}}{\text{. }} & {p_{k,i}} \geqslant \max \left\{ {{{p}_{\text{R}}},{{p}_{\theta }},{{p}_{d}}} \right\} \triangleq {p_{\text{m} ,k}},\forall k, \label{e17}\\ 
	&\sum\limits_{k=1}^K {{p_{k,i}}}  \leqslant {P_{{\text{M}}}}\label{e18},  
\end{align}
where 
\begin{align}
	p _{\rm{R}} &= {{\left[ {\exp \left( {{{{{\cal A}_{k,i}}} \mathord{\left/
							{\vphantom {{{{\cal A}_{k,i}}} {\min \left\{ {\Delta T,{T_{\rm{d}}}} \right\}}}} \right.
							\kern-\nulldelimiterspace} {\min \left\{ {\Delta T,{T_{\rm{d}}}} \right\}}}} \right) - 1} \right]} \mathord{\left/
			{\vphantom {{\left[ {\exp \left( {{{{{\cal A}_{k,i}}} \mathord{\left/
										{\vphantom {{{{\cal A}_{k,i}}} {\min \left\{ {\Delta T,{T_{\rm{d}}}} \right\}}}} \right.
										\kern-\nulldelimiterspace} {\min \left\{ {\Delta T,{T_{\rm{d}}}} \right\}}}} \right) - 1} \right]} {{{\cal B}_{k,i}}}}} \right.
			\kern-\nulldelimiterspace} {{{\cal B}_{k,i}}}},\\ 
p _\theta  &= {\left(1 \mathord{\left/
	 		{\vphantom {1 {{\xi _\theta }}}} \right.
	 		\kern-\nulldelimiterspace} {{\xi _\theta }}\right)}\sum\limits_{m = 1}^4\left( {{{{{\left| {{{ b}_{1m}}} \right|}^2}} \mathord{\left/
	 			{\vphantom {{{{\left| {{{\tilde b}_{1m}}} \right|}^2}} {{\lambda _{m,k,i}}}}} \right.
	 			\kern-\nulldelimiterspace} {{\lambda _{m,k,i}}}}}\right),\\ 
p _d &= {\left(1 \mathord{\left/
			{\vphantom {1 {{\xi _d}}}} \right.
			\kern-\nulldelimiterspace} {{\xi _d}}\right)}\sum\limits_{m = 1}^4 \left( {{{{{\left| {{{ b}_{2m}}} \right|}^2}} \mathord{\left/
				{\vphantom {{{{\left| {{{\tilde b}_{2m}}} \right|}^2}} {{\lambda _{m,k,i}}}}} \right.
				\kern-\nulldelimiterspace} {{\lambda _{m,k,i}}}}}\right). \label{aa1}
\end{align}
\subsection{Low-complexity Power Allocation to ${\cal P}{3}$} \label{sec:low-c}
Observed from (\ref{e17})$\text{-}$(\ref{aa1}), if $\sum\nolimits_{k = 1}^K {{p_{{\rm{m}},k}}}  > {P_{\rm{M}}}$, the optimization problem ${\cal P}{3}$ has no solution. Therefore, we next only consider to the cases where a solution can be found to the optimization problem. Before solving the optimal power allocation scheme for the RSU, we first introduce the following proposition.

\textit{Proposition 1}: Whether for the original optimization problem ${\cal P}{1}$, or for the simplified optimization problem ${\cal P}{3}$, all the vehicles should have the same transmit delay, i.e.,
\begin{equation}
{T_{k,i}} = \frac{{{{\cal A}_{k,i}}}}{{\ln \left( {1 + {{\cal B}_{k,i}}{p_{k,i}}} \right)}} \triangleq T,\forall k \label{e22}.
\end{equation}
\begin{IEEEproof}
We prove \textit{Proposition 1} by contradiction. According to the optimal power allocation, the transmit delay should satisfy
\begin{align}
{T_{a,i}\left({p_{a,i}}\right)} > {T_{k,i}\left({p_{k,i}}\right)} > {T_{b,i}\left({p_{b,i}}\right)},k \ne a,b.
\end{align}
Based on (\ref{aa3}), the transmit delay ${T_{a,i}}\left(  \cdot  \right)$ and ${T_{b,i}}\left(  \cdot  \right)$ are monotonically decreasing functions. Therefore there should exist $\Delta p \in \left( {0,{p_{b,i}}-{p_{\rm{m}}}} \right)$ satisfying that
\begin{align}
	{T_{a,i}\left({p_{a,i}}\right)} &> {T_{a,i}\left({p_{a,i}+\Delta p}\right)} \notag \\
	&= {T_{b,i}\left({p_{b,i}-\Delta p}\right)} > {T_{b,i}\left({p_{b,i}}\right)}.
\end{align}  
Then, we can define the new power allocation ${{\tilde p}_{a,i}} = {p_{a,i}} + \Delta p,{{\tilde p}_{b,i}} = {p_{b,i}} - \Delta p,{\tilde p_{k,i}} = { p_{k,i}}$$(k \ne a,b)$, then the maximum transmit delay decreases, which contradicts to the min-max criterion. Therefore, all the vehicles should have the same transmit delay, which is denoted as $T$.
\end{IEEEproof}
Based on \textit{Proposition 1}, the optimal power allocation scheme of the RSU can be written as
\begin{equation}
{p_{k,i}^{*}} = \frac{1}{{{{\cal B}_{k,i}}}}\left[ {{\text{exp} \left( {\frac{{{{\cal A}_{k,i}}}}{T}}  \right) } - 1} \right].\label{e20}
\end{equation}
Due to the monotonically decreasing property, to minimize the maximum transmit delay among all vehicles, we should have
\begin{equation}
  \sum\limits_{k=1}^K {{p_{k,i}}}  = {P_{{\text{M}}}}. \label{e400}
\end{equation}
Substituting (\ref{e20}) into (\ref{e400}), we can obtain
\begin{equation}
	\sum\limits_{k=1}^K {\frac{1}{{{{\cal B}_{k,i}}}}\left[ {{  \text{exp} \left( {\frac{{{{\cal A}_{k,i}}}}{T}} \right)    } - 1} \right]}  = {P_{{\text{M}}}}.
	\label{e21}
\end{equation}

In terms of (\ref{e21}) and the required information size $D_{k,i}$ for different vehicles, the optimal power allocation ${p_{k,i}^{*}}$ in (\ref{e20}) can be discussed in two cases. 
\begin{itemize}
	\item \textbf{Case 1} (${{D}_{k,i}} = {D}$). When the size of information $D_{k,i}$ is the same for all vehicles, we have ${\cal A}_{k,i}={\cal A}$ based on (\ref{aa2}). Then, based on (\ref{e21}), we have
	\begin{equation}
	T = {\cal A} {\left\{ {{\ln \left[ {1 + {P_{{\text{M}}}}{{\left( {\sum\limits_{k=1}^K {\frac{1}{{{{\cal B}_{k,i}}}}} } \right)}^{ - 1}}} \right]}} \right\}}^{-1} .\label{e31}
	\end{equation}
	Substituting (\ref{e31}) into (\ref{e20}) results in the optimal power allocation, i.e.,
	\begin{equation}
	{p_{k,i}^ *} = {{{P_{{\text{M}}}}}} { \left( {{{{\cal B}_{k,i}}\sum\limits_{k=1}^K {\frac{1}{{{{\cal B}_{k,i}}}}} }}\right)  }^{-1}. \label{optimal solving}
	\end{equation}
	\item \textbf{Case 2} (${{D}_{k,i}} \ne {D}$). When the size of information $D_{k,i}$ is different for all vehicles, we have ${\cal A}_{k,i} \ne {\cal A}$ based on (\ref{aa2}). In order to minimize the maximum transmit delay among vehicles, the power allocation of the RSU can be obtained by designed Alg. 1 based on bi-section search. The complexity of Alg. 1 contains two parts, namely a sorting algorithm (line 2) and a loop (line 3). The sorting algorithm has a complexity of ${\cal O}\left( K\right)$. On the other hand, upon denoting the number of iterations by $T$ and noting that ${{\left( {{{\max  \limits_k}}\left\{ {{{{D_{k,i}}} \mathord{\left/ {\vphantom {{{D_{k,i}}} {{R_{k,i}}}}} \right. \kern-\nulldelimiterspace} {{R_{k,i}}}}} \right\} - {{\min  \limits_k}}\left\{ {{{{D_{k,i}}} \mathord{\left/	{\vphantom {{{D_{k,i}}} {{R_{k,i}}}}} \right. \kern-\nulldelimiterspace} {{R_{k,i}}}}} \right\}} \right)} \mathord{\left/		{\vphantom {{\left( {{{\max }_{k \in K}}\left\{ {{{{D_{k,i}}} \mathord{\left/	{\vphantom {{{D_{k,i}}} {{R_{k,i}}}}} \right. \kern-\nulldelimiterspace} {{R_{k,i}}}}} \right\} - {{\min }_{k \in K}}\left\{ {{{{D_{k,i}}} \mathord{\left/ {\vphantom {{{D_{k,i}}} {{R_{k,i}}}}} \right. \kern-\nulldelimiterspace} {{R_{k,i}}}}} \right\}} \right)} {{2^T}}}} \right. \kern-\nulldelimiterspace} {{2^T}}} \leq {\varepsilon _{\rm{T}}}$ is a sufficient condition for the loop to stop based on the termination condition ${T_{\rm{U}}} - {T_{\rm{L}}} \leq {\varepsilon _{\rm{T}}}$, the complexity of Alg. 1 can be formulated as 
		\begin{align}
			&{\cal O}\left( {\max \left\{ {{K},{{\log }_2}\left( {{{{M_t}} \mathord{\left/
								{\vphantom {{M} {{\varepsilon _{\rm{T}}}}}} \right.
								\kern-\nulldelimiterspace} {{\varepsilon _{\rm{T}}}}}} \right)} \right\}} \right), \notag \\
			&\quad {M_t} = {\max \limits_k }\left\{ {{{{D_{k,i}}} \mathord{\left/	{\vphantom {{{D_{k,i}}} {{R_{k,i}}}}} \right.	\kern-\nulldelimiterspace} {{R_{k,i}}}}} \right\} - {\min \limits_k}\left\{ {{{{D_{k,i}}} \mathord{\left/	{\vphantom {{{D_{k,i}}} {{R_{k,i}}}}} \right. \kern-\nulldelimiterspace} {{R_{k,i}}}}} \right\}.
	\end{align}	

\end{itemize}
\subsection{Optimal Power Allocation to ${\cal P}{1}$}

The optimal power allocation ${p_{k,i}^ *}$ in Section~\ref{sec:low-c} is obtained by relaxing the constrains (\ref{e12})$\text{-}$(\ref{e13}), therefore it may not be the optimal solution for the original problem ${\cal P}{1}$. To further improve the system performance, we design Alg. 2, based on Alg. 1 for optimal power allocation, but not a closed-formed expression. Based on \textit{Proposition 1}, we can know that the transmit delay should be equal among all $K$ vehicles. On the other hand, $T_{k,i}$ is a monotonically decreasing function with respect to $p_{k,i}$. Therefore, we can increase the power for the vehicle with maximum transmit delay, and correspondingly reduce the same power for the vehicle with the minimum transmit delay, until the delay difference among vehicles vanishes. Based on above discussion, Alg. 2 is proposed to achieve the optimal solution $p_{k,i}^\circ $ for the original problem, where line 7 is used to avoid the ping-pong phenomenon among the vehicles with maximum and minimum transmit delays. Similar to Alg. 1, the complexity of Alg. 2 is also determined by two components, a loop (line 3) and a sort algorithm (line 4). Hence, we can express the total complexity of Alg. 2	as ${\cal O}\left( {{K} \times {{\log }_2}\left( {{{\Delta p} \mathord{\left/{\vphantom {{\Delta p} {{\varepsilon _{\rm{T}}}}}} \right.	\kern-\nulldelimiterspace} {{\varepsilon _{\rm{p}}}}}} \right)} \right)$, which is more complex than Alg. 1. Therefore, we only adopt Alg. 2 when Alg. 1 is unavailable in the low transmit power at the RSU. In the simulation sections, we will discuss the applicable conditions of Alg.1 and Alg. 2 in details.

\begin{figure}[!t]
	\label{alg:LSB}
	\renewcommand{\algorithmicrequire}{\textbf{Input:}}
	\renewcommand{\algorithmicensure}{\textbf{Output:}}
	\removelatexerror
	\begin{algorithm}[H]
		\caption{\textbf{:}\; Delay Iterative Optimization Algorithm}
		\begin{algorithmic}[1]
			\STATE Initialize the threshold $\varepsilon_\text{T} $ and ${p_{k,i}}=\frac{1}{K}{P_{{\text{M}}}}$.
			\STATE Calculate $T_{k,i}$ based on (\ref{e30}) and compute ${T_\text{L}}=\min \left\{ {{T_{k,i}}} \right\},{T_\text{U}}=\max \left\{ {{T_{k,i}}} \right\},\forall k$.
			\WHILE{${T_\text{U}} - {T_\text{L}}>{\varepsilon_\text{T}}$}
			\STATE Let ${T} = \frac{1}{2}\left( {{T_\text{L}} + {T_\text{U}}} \right)$.
			\STATE If $\sum\nolimits_{k=1}^K {\frac{1}{{{{\cal B}_{k,i}}}}\left[ {{  \text{exp} \left( {\frac{{{{\cal A}_{k,i}}}}{T_0}} \right)    } - 1} \right]}>{P_{{\text{M}}}}$, let ${T_\text{L}} = {T}$; Else ${T_\text{U}}={T}$.
			\ENDWHILE
			\STATE Output ${p_{k,i}^ *}$ according to (\ref{e20}).
		\end{algorithmic}
	\end{algorithm}
	\vspace{-0.15cm} 
\end{figure}
\begin{figure}[!t]
	\label{alg:LSB}
	\renewcommand{\algorithmicrequire}{\textbf{Input:}}
	\renewcommand{\algorithmicensure}{\textbf{Output:}}
	\removelatexerror
	\vspace{-0.15cm} 
	\begin{algorithm}[H]
		\caption{\textbf{:}\; Complementary Iterative Algorithm}
		\begin{algorithmic}[1]
			\STATE Initialize the threshold ${\varepsilon_\text{p}}$, $\Delta p$ and ${p_{k,i}}=\frac{1}{K}{P_{{\text{M}}}}$.
			\STATE Calculate $T_{k,i}$ based on (\ref{e30}).
			\WHILE{${\Delta p} > {\varepsilon_\text{p}}$}
			\STATE Let $\tilde k_{\text{M}} = \arg \mathop {\max }\limits_k \left\{ {{T_{k,i}}} \right\}$ and $\tilde k_{\text{m}} = \arg \mathop {\min }\limits_k \left\{ {{T_{k,i}}} \right\}$.
			\STATE Let $p_{{\tilde k_{\text{M}}},i}=p_{{\tilde k_{\text{M}}},i}+\Delta p$ and $p_{{\tilde k_{\text{m}}},i}=p_{{\tilde k_{\text{m}}},i}-\Delta p$.
			\STATE Calculate $T_{{\tilde k_{\text{M}}},i}$ and $T_{{\tilde k_{\text{m}}},i}$ based on (\ref{e30}).
			\STATE If $T_{{\tilde k_{\text{M}}},i} < T_{{\tilde k_{\text{m}}},i}$, let $p_{{\tilde k_{\text{M}}},i}=p_{{\tilde k_{\text{M}}},i}-\Delta p$, $p_{{\tilde k_{\text{m}}},i}=p_{{\tilde k_{\text{m}}},i}+\Delta p$ and ${\Delta p} = {\Delta p} / 2$.
			\ENDWHILE
			\STATE Output ${p_{k,i}^\circ} = {p_{k,i}}$.
		\end{algorithmic}
	\end{algorithm}
\end{figure}
\section{Simulation Results and Discussions}
\subsection{Parameter Setup}
In the simulation, the default parameters are given as follows. The distance between the RSU and the center of the road is $H = 4$ $\text{m}$. The carrier frequency of the signal, the system bandwidth and the slot interval are $f_\text{c} = 30\text{ GHz}$, $B = 400\text{ MHz}$ and $\Delta T = 0.01\text{ s}$, respectively. The noise power at the RSU and vehicles are $\sigma _\text{R}^2 = \sigma _\text{C}^2 = 0.0025$.        

%
\subsection{Transmit Delay Results}
 Fig. 2 illustrates the transmit delay versus the number of iterations for the Alg. 1. When the number of iterations increases, the lower bound $T_\text{L}$ increases and the upper bound $T_\text{U}$ decreases, and they rapidly converge to a constant, which demonstrates the convergence of Alg. 1. On the other hand, the dashed lines in Fig. 4 show the transmit delay versus the number of vehicles for Alg. 1. With the increasing number of vehicles, the convergence value of Alg. 1 increases slowly, because the amount of power allocated to each vehicle decreases.

\begin{figure}[!t]
	\includegraphics[width=3.5in]{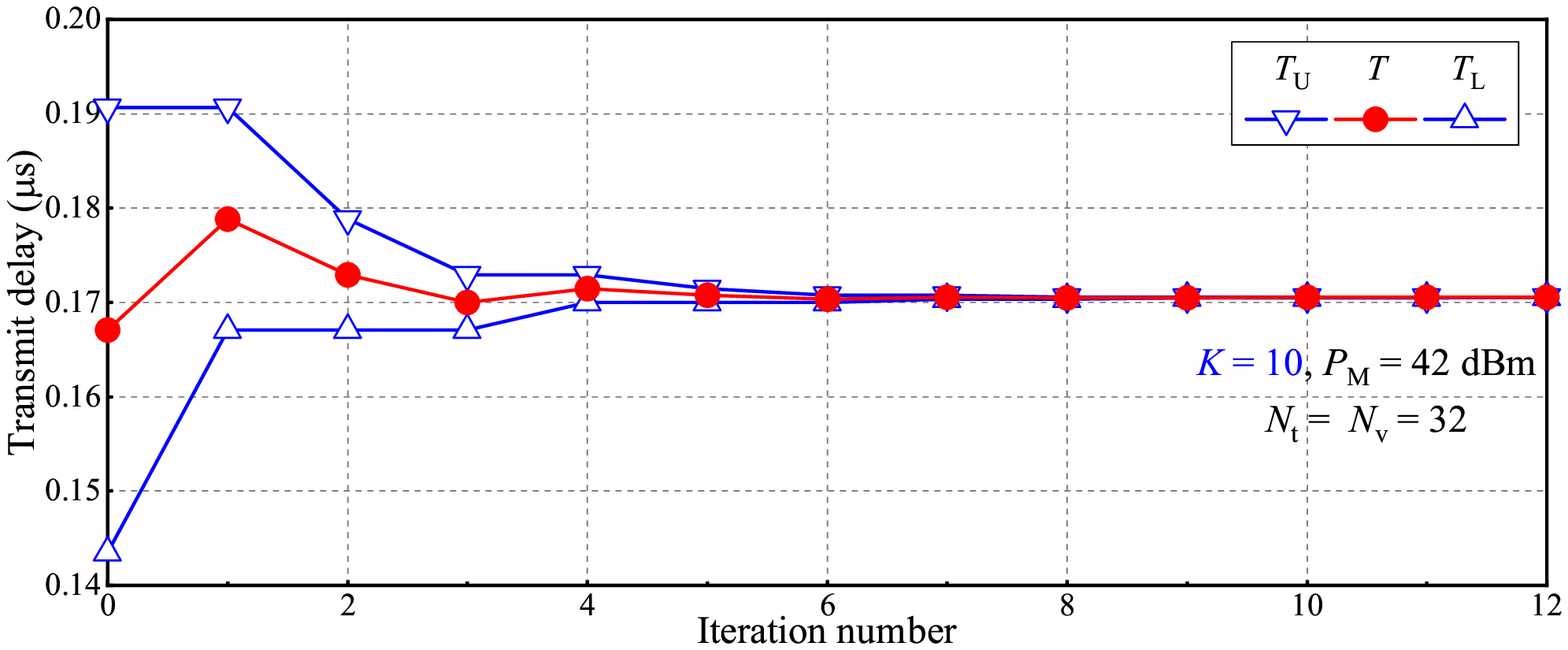}
	\caption{Convergence of Alg. 1.}
\end{figure}
\begin{figure}[t]
	\centering
	\includegraphics[width=3.5in]{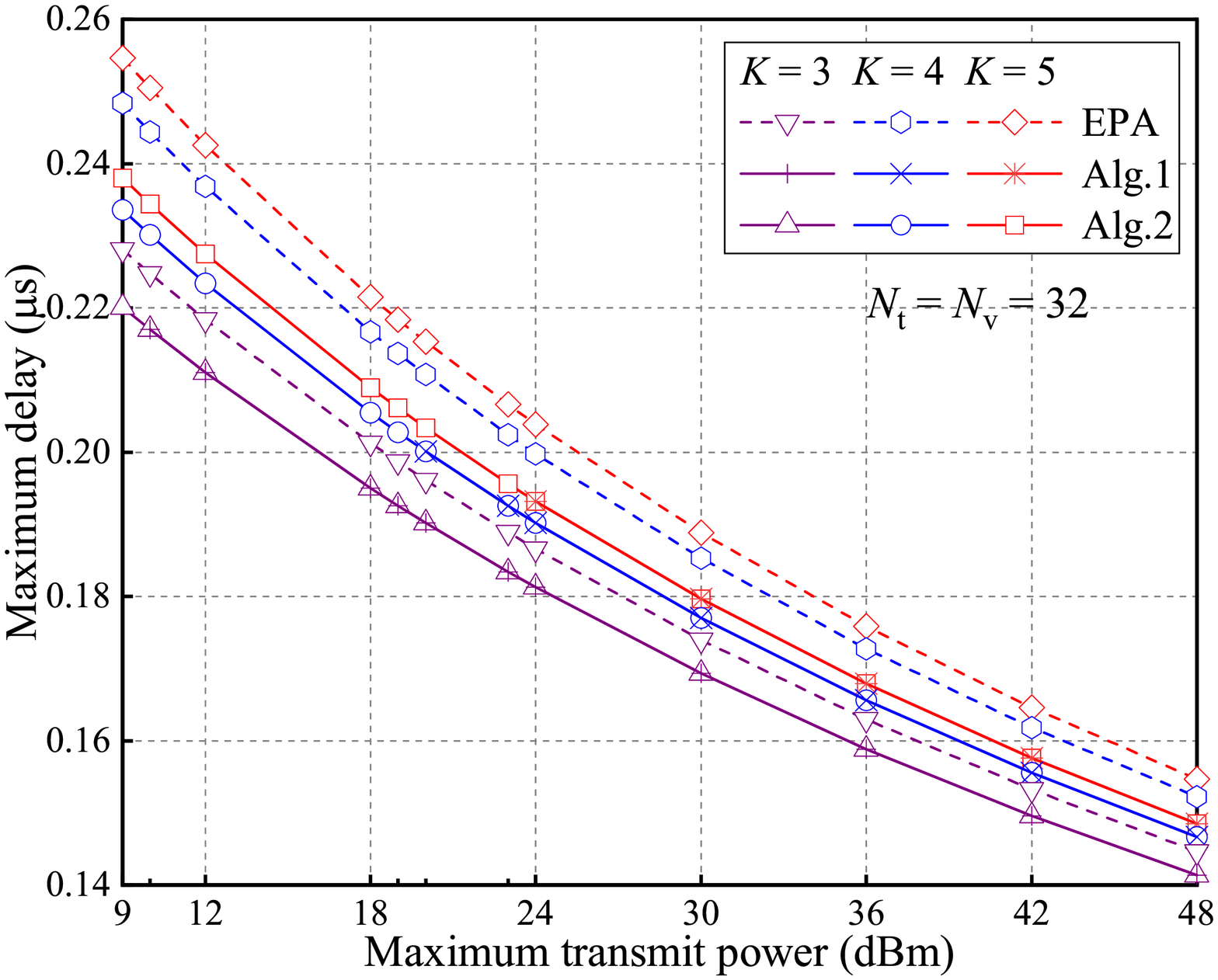}
	\caption{Maximum delay versus maximum transmit power at the RSU.}
	\label{3}
\end{figure}
\begin{figure}[t]
	\centering
	\includegraphics[width=3.5in]{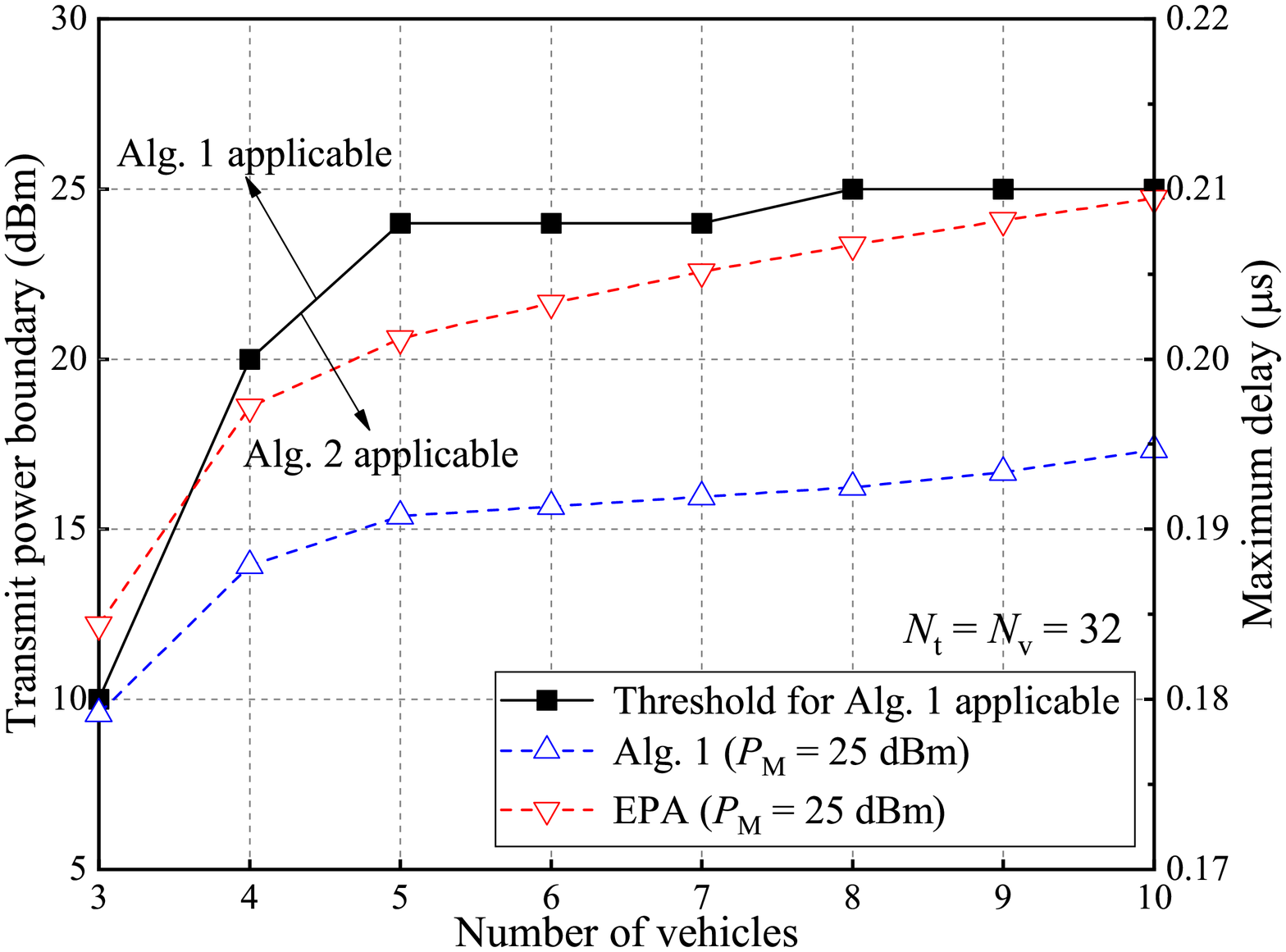}
	\caption{The transmit power boundary between Alg. 1 and Alg. 2 and the maximum delay versus the number of vehicles.}
	\label{3}
\end{figure}
\begin{figure}[t]
	\centering
	\includegraphics[width=3.5in]{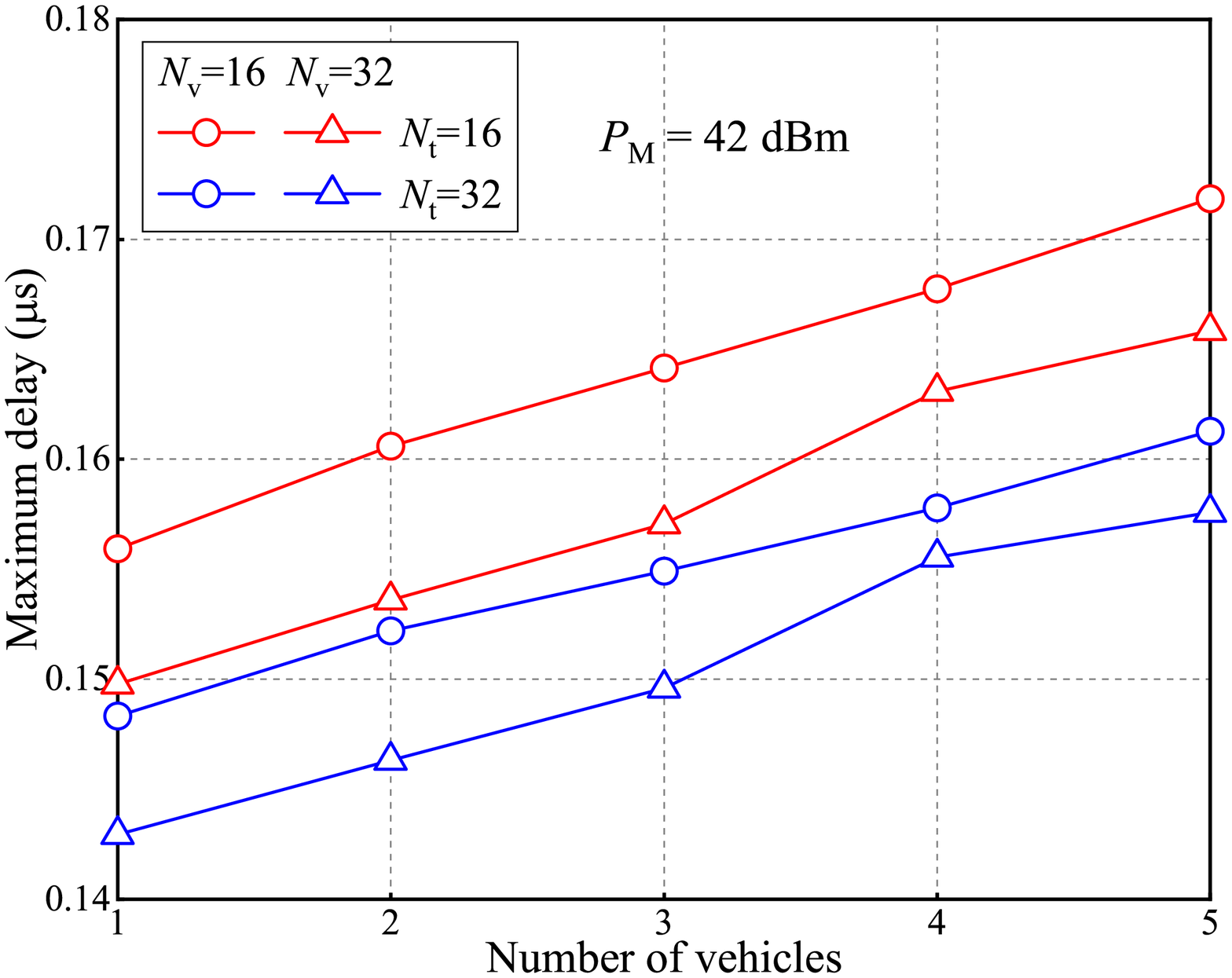}
	\caption{Maximum delay versus the number of vehicles.}
	\label{4}
\end{figure}

	

The maximum delay versus the maximum transmit power is illustrated in Fig. 3. With low transmit power at the RSU, there may be no solution for the formulated problem based on Alg. 1; however Alg. 2 can still obtain the optimal power allocation to minimize the maximum delay. When the maximum transmit power becomes high, the relaxing operation in constrains (\ref{e12})$\text{-}$(\ref{e13}) can be omitted for Alg. 1, however it becomes fatal for Alg. 1 under the limited transmit power. Furthermore, with the increasing number of vehicles, the applicable power threshold of Alg. 1 increases because the allocated power of each vehicle decreases. Moreover, both the proposed algorithms outperform the equal power allocation (EPA) scheme~\cite{a27} due to the reasonable power allocation among vehicles.

Fig. 4 demonstrates both the transmit power boundary between Alg. 1 and Alg. 2 and the maximum delay versus the number of vehicles. Compared with Fig. 3, the solid line with square symbol in Fig. 4 further illustrates the minimum transmit power requirement for Alg. 1. With the increasing number of vehicles, the transmit power boundary tends to increase because the increasing vehicles require more power to guarantee the minimum performance requirements. In addition, the two dashed lines in Fig. 4 show an upward trend and the gap between the two dashed lines increases with the increasing number of vehicles. The increasing gap indicates that larger gain could be achieved by our proposed algorithms compared with EPA.
	

Fig. 5 plots the maximum delay versus the number of vehicles. With the increasing number of antennas, the transmit delay decreases, for a given number of vehicles. On the other hand, increasing the number of transmit antennas is more effective to reduce the transmit delay than increasing the number of receive antennas, which can be observed by the red lines with triangles and the blue lines with circles. The main reason is that the receive beams are calculated by the predicted angles of two-slot, while the transmit beams only need the predicted angles of one-slot, which results in a smaller error.

\section{Conclusions}

The DFRC signal could cut the uplink communication overhead and effectively reduce the signal transmit delay in IoV. This paper proposed a scheme for transmit power allocation at the RSU to minimize the maximum transmit delay among vehicles, taking into account the PSI constraints. After problem analysis and transformation, two iterative algorithms were proposed with different computational complexities and transmit power levels. Simulation results indicated that the proposed algorithms could significantly reduce the maximum transmit delay among vehicles. And increasing the number of RSU transmit antennas was more effective than increasing the number of vehicle's receive antennas. DFRC-based power allocation for minimizing communication transmit delay is an exciting and emerging area of research. While this paper mainly focuses on integrating radar and communication functions for straight-line road scenario due to the page limit, future research challenges and directions still need to be explored, such as the optimization problems of full-duplex DFRC, and unmanned aerial vehicles (UAVs) aided DFRC.


\end{document}